\documentclass[preprint]{revtex4-1}
\usepackage[utf8]{inputenc}
\usepackage[english]{babel}
\usepackage{amssymb,amsmath}
\usepackage{siunitx}
\usepackage{graphicx}
\usepackage{hyperref}

\newcommand{\Berd}{\mathbf{B}_{0}}
\newcommand{\Berdn}{\left\vert\mathbf{B}_{0}\right\vert}
\newcommand{\Berda}{B_{0}}
\newcommand{\Bpp}{\mathbf{B}_{\mathrm{p}}}
\newcommand{\Bppn}{\left\vert\mathbf{B}_{\mathrm{p}}\right\vert}
\newcommand{\Bppa}{B_{\mathrm{p}}}
\newcommand{\Btotal}{\mathbf{B}}
\newcommand{\Bstar}{B^{\ast}}
\newcommand{\Btx}{\mathbf{B_{\mathrm{T}}}}
\newcommand{\Beff}{\mathbf{B_{\mathrm{eff}}}}
\newcommand{\Btp}{\mathbf{B}_{\mathrm{T}}^{+}}
\newcommand{\Brm}{\mathbf{B}_{\mathrm{R}}^{-}}
\newcommand{\bt}{\mathbf{b}_{\mathrm{T}}^{\perp}}
\newcommand{\br}{\mathbf{b}_{\mathrm{R}}^{\perp}}

\newcommand{\Mtotal}{\mathbf{M}}
\newcommand{\Mnull}{\mathbf{M}_0}
\newcommand{\Mnulla}{M_0}
\newcommand{\Mpp}{\mathbf{M_{\mathrm{p}}}}
\newcommand{\Mppn}{\left\vert\mathbf{M_{\mathrm{p}}}\right\vert}
\newcommand{\Mppa}{M_{\mathrm{p}}}

\newcommand{\Mperp}{M_{\perp}}

\newcommand{\gammaH}{\gamma_{\mathrm{H}}}
\newcommand{\larmor}{\omega_{\mathrm{L}}}
\newcommand{\larmorf}{f_{\mathrm{L}}}
\newcommand{\larmorb}{2\pi f_{\mathrm{L}}}

\newcommand{\Iac}{I_{\mathrm{ac}}}
\newcommand{\Ipp}{I_{\mathrm{p}}}
\newcommand{\tauramp}{\tau_{\mathrm{r}}}
\newcommand{\tauwait}{\tau_{\mathrm{w}}}
\newcommand{\taudead}{\tau_{\mathrm{d}}}

\newcommand{\myexp}{EXP}
\newcommand{\mylinexp}{LINEXP}
\newcommand{\myhalfcos}{HCOS}
\newcommand{\mylin}{LIN}

\begin{document}
	
\title{Utilizing pre-polarization to enhance SNMR signals -- effect of imperfect switch-off}	
	
\author{Hiller, Thomas}
\email[Corresponding author: ]{thomas.hiller@leibniz-liag.de}
\author{Dlugosch, Raphael}
\author{M\"uller-Petke, Mike}
\affiliation{Leibniz Institute for Applied Geophysics, Stilleweg 2, 30652 Hannover, Germany}

\begin{abstract}
Surface nuclear magnetic resonance (SNMR) is a well-established technique for the hydrogeological characterization of the subsurface up to depths of about \SI{150}{\meter}. Recently, SNMR has been adapted to investigate also the shallow unsaturated zone with small surface loop setups. Due to the decreased volume, a pre-polarization (PP) field prior to the classical spin excitation is applied to enhance the measured response signal. Depending on the strength and orientation of the applied PP-field, the enhancement can often reach several orders of magnitude in the vicinity of the PP-loop. The theoretically achievable enhancement depends on the assumption of an adiabatic, i.e. perfect, switch-off of the corresponding PP-field. To study the effect of imperfect switch-off, we incorporate full spin dynamics simulations into the SNMR forward modeling. The affected subsurface volume strongly depends on the chosen PP switch-off ramp and the geometry of the loop setup. Due to the imperfect switch-off, the resulting SNMR sounding curves can have significantly decreased signal amplitudes. For comparison,  the signal amplitudes of either a \SI{1}{\milli\second} exponential or linear switch-off ramp are reduced by \SI{17}{\percent} and \SI{65}{\percent}, respectively. Disregarding this effect would therefore yield an underestimation of the corresponding subsurface water content of similar magnitude.
\end{abstract}


\date{01.05.2020}

\thanks{This preprint was accepted in Geophys. J. Int.}
\preprint{{\textit{Geophys. J. Int.}} (0000) \textbf{000}, 000–000}

\maketitle

\section{Introduction}
\label{sec:intro}

Surface nuclear magnetic resonance (SNMR), the surface-based variant of nuclear magnetic resonance (NMR), is one of the ``younger'' methods employed in applied hydrogeophysics. It has evolved over the last three decades to a reliable and well-established method to characterize near surface aquifer systems \cite[e.g.][]{legchenko2004GW,vouillamoz2011NSG,behroozmand2015SiG}. SNMR is the only non-destructive surface method that allows a direct depth-resolved quantification of the water content of such aquifer systems. Furthermore, the measured signal also carries information about the corresponding pore space and therefore can be used to establish relationships between the porous medium and its hydraulic properties \cite[e.g.][]{seevers1966TS,mohnke2008JoAG,dlugosch2013G}. Over the years, SNMR underwent several iterations of improvement in terms of hardware development, signal processing and forward / inverse modeling. It is successfully in use as a magnetic resonance tomography (MRT) tool to investigate 2D and 3D subsurface structures \cite[e.g.][]{hertrich2007IToGaRS,hertrich2008PiNMRS,legchenko2011NJoP,dlugosch2014NSG,jiang2016G,jiang2018G}. By now, it is also commonly applied together with other geophysical techniques like vertical electrical sounding (VES), electrical resistivity tomography (ERT) or ground penetrating radar (GPR) to increase structural model resolution and reduce the ambiguity of a single method \cite[e.g.][]{vouillamoz2007JoAG,gunther2012HaESS,costabel2017JoAG,jiang2017eage,skibbe2018G}.

Because of the weak NMR signal strengths involved, SNMR is generally prone to electromagnetic noise. When targeting the shallow unsaturated zone with small loop layouts, the noise sensitivity is even more troublesome due to the even weaker NMR signals. To handle electromagnetic noise, several efficient noise cancellation techniques have been developed \cite[e.g.][]{costabel2014NSG,larsen2014GJI,mueller-petke2014NSG}. A comprehensive overview on SNMR post-processing techniques can be found in \cite{mueller-petke2016G}. Besides the use of noise cancellation techniques, there are other approaches available to increase the signal-to-noise ratio (SNR) of SNMR measurements. One is the application of adiabatic fast passage pulses (AP), where the amplitude and frequency of the excitation pulse is modulated in a time short compared to the relaxation times \cite[e.g.][]{powles1958PPS,tannus1997NiB,grunewald2016G,grombacher2018G}. The term \emph{adiabatic} describes the rotational motion of the magnetization vector in relation to the varying magnetic field. The closer the magnetization is ``locked'' to the magnetic field the more adiabatic is the process. In the presence of inhomogeneous excitation fields (general case when using surface loops), this yields a spatially more homogeneous excitation and therefore enhanced signal amplitudes compared to the standard on-resonance (OR) excitation which operates at the resonance frequency. AP-excitation has its origins in other fields of NMR and is applied in medical imaging and microscopy \cite[e.g.][]{ugurbil1988JMR,zijl1996JAmCS}. However, until today OR-excitation is by far the most common excitation technique used in SNMR, whereas AP-excitation has been introduced to SNMR just within the past few years \cite[]{grunewald2016G}.

Another technique to increase the SNR of SNMR measurements is the pre-polarization (PP) of the net magnetization by applying an enhanced background static magnetic field prior to performing the actual measurement at a weaker field \cite[]{packard1954PR}. Therefore, it is also possible to conduct NMR measurements in the weak Earth's magnetic field if a proper pre-polarization is applied \cite[e.g.][]{melton1971RSI,planinsic1994JMRA,callaghan1997RoSI}. The effect of pre-polarization to SNMR measurements (SNMR-PP) with small loop setups ($d_{\mathrm{pp}}\approx\SI{5}{\meter}$) was conceptually shown by \cite{pasquale2014VZJ} for targets in the shallow vadose zone. The first field SNMR-PP measurement was conducted on a water reservoir by \cite{lin2017GJI} with a \SI{2}{\meter} square PP-loop and \SI{1200}{\ampere} effective DC. Quite recently, the application of SNMR-PP for underground tunnel measurements has also attracted interest within the SNMR community \cite[]{lin2019IEEE,costabel2019JAG}. Furthermore, although not being a classical hydrogeophysical SNMR application in terms of hydro\-geophysical aquifer charac\-teri\-zation, the detection of oil under sea ice from a mobile, helicopter-borne NMR device also makes use of pre-polarization techniques to amplify the measured NMR signal \cite[]{conradi2018JoMRa,altobelli2019MPB}.

\begin{figure}
	\centering
	\includegraphics[width=1\linewidth]{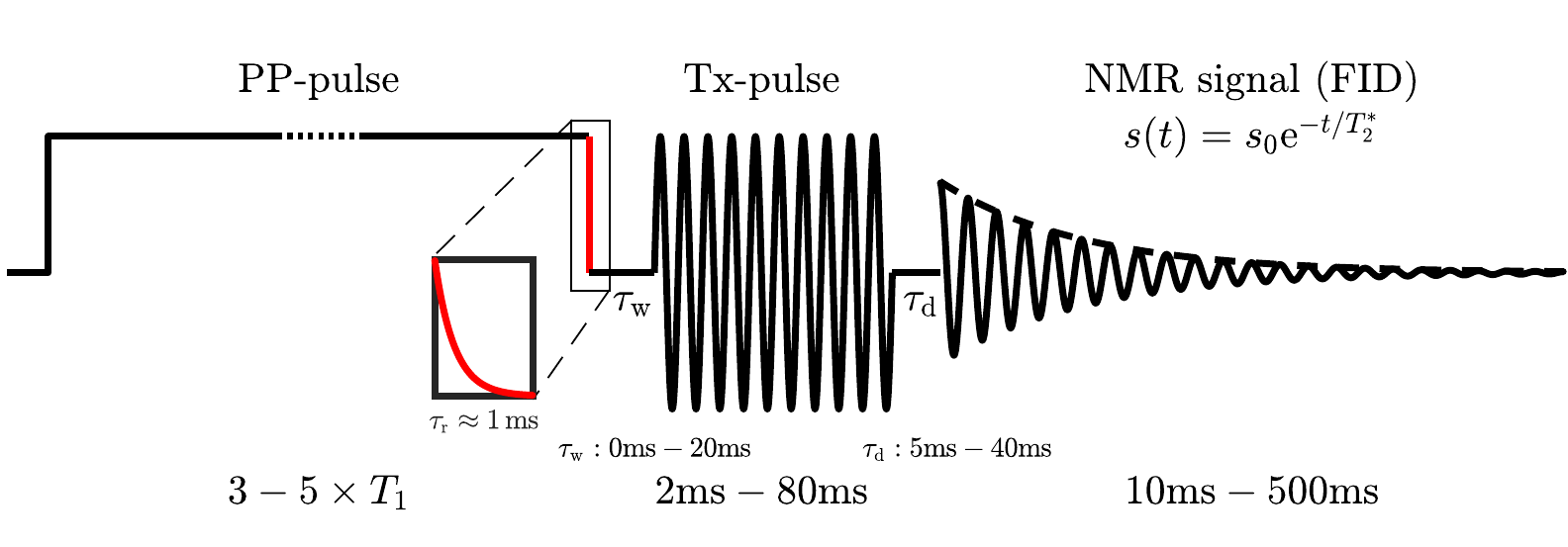}
	\caption{Sketch showing a typical SNMR-PP measurement sequence: PP-pulse with switch-off ramp (red) $\rightarrow$ wait time $\tauwait$ $\rightarrow$ Tx-pulse (on-resonant or adiabatic) $\rightarrow$ dead time $\taudead$ $\rightarrow$ NMR signal response; note that the axes are not to scale.}
	\label{fig:fig1}
\end{figure}

A typical SNMR-PP measurement sequence is sketched in Fig.~\ref{fig:fig1}. Initially a strong PP $\Bpp$-field is generated by energizing a PP-loop with a large DC current $\Ipp$ for a duration long enough to polarize the spins to a known level to enhance the subsequent signal amplitude. The length of the PP-pulse (\si{\milli\second} to \si{\second}) depends on the $T_1$ relaxation time of the investigated medium as the enhanced magnetization builds up with $T_1$. After switching-off the PP-field within the switch-off ramp time $\tauramp$, a pulse (Tx) is used to excite the hydrogen protons in the subsurface. Subsequently a NMR response, typically a free induction decay (FID), can be recorded for up to several hundreds of milliseconds depending on the particular relaxation times of the investigation target. By increasing either the length or amplitude of the Tx-pulse, it is possible to excite deeper and larger volumes in the subsurface and therefore gather quantitative, depth-resolved information about water content and relaxation time.

Common to the aforementioned SNMR-PP applications, is the assumption that the PP switch-off itself is carried out properly. This means that the PP-field has to be switched-off (or better ramped down - red line in Fig.~\ref{fig:fig1}) in a particular manner during the time span $\tauramp$, so that the enhanced magnetization is perfectly aligned with the Earth's magnetic field at the end of the switch-off. Two important things need to be considered here. Due to its spatial inhomogeneity, the magnitudes of the PP-field range over several orders of magnitude. Furthermore, the spatial distribution of relative orientations between the Earth's magnetic field and the PP-field varies with the inclination of the Earth's magnetic field and locally ranges between \ang{0} and \ang{180}. For practical considerations this means that a particular switch-off ramp (with a fixed switch-off time $\tauramp$) employed in a certain SNMR-PP device, needs to be optimized for all combinations of PP-field amplitudes and relative orientations. Considering this rather large parameter space it is very unlikely that the maximal theoretically possible enhanced magnetization is established over the entire affected subsurface after the PP switch-off \cite[e.g.][]{melton1995JoMRSA,melton2002JoMR,conradi2017JoMR}. Furthermore, when targeting the shallow subsurface (e.g. vadose zone), which features $T_1$ relaxation times even shorter than about \SI{100}{\milli\second}, the time span between PP switch-off and FID record should be as short as possible to avoid signal loss due to $T_1$ relaxation. This time span consists of wait time $\tauwait$, Tx-pulse excitation and dead time $\taudead$ (cf.~Fig.~\ref{fig:fig1}), where $\tauwait$ and $\taudead$ are generally device-dependent and the Tx-pulse length depends on the measurement protocol and / or target depth. Depending on these parameters it may be possible to prolong the switch-off time $\tauramp$ to increase the adiabatic switch-off performance. In any case, one needs to evaluate the trade-off between switch-off performance due to longer $\tauramp$ and magnetization decrease due to short $T_1$ relaxation.

The objective of this work is to quantify the effect of an imperfect PP switch-off on the resulting NMR signal due to the application of different PP switch-off ramps. To do so, we employ full spin dynamics simulations of the PP-switch-off within the calculation of the three-dimensional SNMR forward kernel. This means that for every point in the subsurface where the $\mathbf{B}$-field is calculated, we need to solve a set of differential equations to determine the magnetization after the PP switch-off. Considering the $\mathbf{B}$-field grid discretization ($\mathrm{nr}\times\mathrm{n\varphi}\times \mathrm{nz} = 71\times361\times192$) later used in section~\ref{sec:theoMRS}, the number of points, and hence the number of needed calculations is almost 5 million. If for a given SNMR-PP setup and PP switch-off ramp, the effect of the PP switch-off on the SNMR signal is negligible, one could omit this extensive and time consuming procedure for any further SNMR kernel calculations. The surface loop sizes exemplarily used in this work are $d=\SI{2}{\meter}$ and are comparable to the sizes used in the works of \cite{lin2017GJI} and \cite{costabel2019NSG}. Consequently, this size roughly holds for any geophysical SNMR-PP setup currently in use and is constrained by the technically applicable DC current that drives the PP-loop (which is in the order of \SI{20}{\ampere} to \SI{100}{\ampere} per turn and an effective DC of about \SI{1000}{\ampere}). If available, we utilize PP switch-off ramp shapes that have been applied by other groups either for geophysical SNMR-PP \cite[]{lin2017GJI,costabel2019NSG} or alternative applications that also employ PP for NMR signal enhancement \cite[]{melton1995JoMRSA,conradi2017JoMR}. The intention of this work is not to present or propose a particularly optimized switch-off ramp for a certain special application. In contrast, we want to raise the awareness that the non-consideration of the PP switch-off might lead to unexpected and undesired results.

Our study was partially inspired by the works of \cite{melton1995JoMRSA,melton2002JoMR,conradi2017JoMR} and uses some of their results as benchmark cases for our numerical implementation (cf.~appendix~\ref{ssec:modeling}). We like to point out that the application of SNMR-PP significantly differs from the works of the aforementioned groups. While \cite{melton1971RSI,melton1995JoMRSA} and \cite{melton2002JoMR} were using NMR with PP in the Earth's magnetic field to study the relaxation of liquids similar to laboratory NMR relaxometry measurements, \cite{conradi2018JoMRa} and \cite{altobelli2019MPB} use NMR with PP to qualitatively detect oil on water under a layer of sea ice. Essentially, and in contrast to SNMR, the latter application is a detection method with additional quantification capability. Because SNMR applications always aim to exactly quantify the amount of water in a particular depth-resolved subsurface volume, they need to employ the extensive forward modeling described above.

This paper is structured as follows: first, we introduce the basic concepts of SNMR including the difference of OR- and AP-excitation. Then, we show the spatial sensitivity of an ideal enhancement of the SNMR signal due to a perfect PP-pulse switch-off. Subsequently, we focus on the imperfect PP-pulse switch-off and carry out a parameter study to evaluate the dependence of the PP-effectiveness of different switch-off ramps on the control parameters. We finalize the results part by comparing the influence of different switch-off ramps on forward modeled SNMR signal amplitudes.

\section{Theoretical basics of SNMR}
\label{sec:theoMRS}

The nuclear spins of hydrogen protons precess around the Earth's magnetic field $\Berd$ (with amplitude $\Berda=\Berdn$) with the angular Larmor frequency $\larmor = \larmorb = -\gammaH\Berda$, where $\gammaH$ is the gyromagnetic ratio of hydrogen. Note, that the negative sign of the resulting Larmor frequency defines only the direction of precession of the spin and is most of the time safely ignored for common SNMR applications. The spins preferentially align with the axis of $\Berd$ and depending on the magnitude of $\Berd$, this alignment generates a small magnetic moment $\Mnull\propto\Berd$ \cite[e.g.][]{levitt2002}. In case of SNMR-PP the magnetic moment $\Mnull$ is proportional to the resulting field $\Btotal=\Berd+\Bpp$, where $\Bpp$ is the magnetic field generated by the PP-loop. In order to stimulate a detectable signal, an electromagnetic Tx-pulse at the Larmor frequency is transmitted through a surface loop to excite the proton spins out of their equilibrium state. Thereby, also $\Mnull$ is forced out of its equilibrium orientation by the excitation magnetic field $\Btx$ into an excited state $\Mtotal$. After the Tx-pulse is switched-off, the spins, and hence also $\Mtotal$, reorient towards their equilibrium orientation parallel to $\Berd$. During this relaxation process $\Mperp$, the component of $\Mtotal$ that is perpendicular to $\Berd$, precesses around $\Berd$ at the Larmor frequency and therewith induces a measurable voltage response in a surface receiver (Rx) loop. The received signal decays exponentially with
\begin{equation}
s(q,t) = s_0(q)\mathrm{e}^{-t/T_2^*}\cos(\larmor t+\phi_s)\;,
\label{eq:signal}
\end{equation}
where $s_0$ is the initial FID amplitude, $T_2^*$ is the free induction decay time and $\phi_s$ is the phase of the signal with respect to the transmitted Tx-pulse. By varying the pulse moment $q=\Iac\tau$, with Tx-pulse current amplitude $\Iac$ and pulse length $\tau$, a 1D sounding curve can be obtained. The measured amplitudes and relaxation times can be inverted for depth-resolved subsurface water content and pore size information, respectively \cite[e.g.][]{legchenko2002JoAG,mueller-petke2010G,mueller-petke2016G}. The forward operator corresponding to eq.~\eqref{eq:signal} is \cite[]{hertrich2008PiNMRS}
\begin{equation}
s_0(q) = \int K(q,\mathbf{r})f(\mathbf{r})d\mathbf{r}\;,
\label{eq:forward}
\end{equation}
with water content distribution $f(\mathbf{r})$ and sensitivity kernel $K(q,\mathbf{r})$. For the general case of separated Tx- and Rx-loops $K(q,\mathbf{r})$ is given by
\begin{align}
\begin{split}
K(q,\mathbf{r}) = &2\larmor M_0\sin\left(-\gamma q \left|\Btp(\mathbf{r})\right|\right)\\&\times\left|\Brm(\mathbf{r})\right|\cdot\mathrm{e}^{i\left[\zeta_{T}(\mathbf{r})+\zeta_{R}(\mathbf{r}
)\right]}\\ &\times \left[\br(\mathbf{r})\cdot\bt(\mathbf{r})+i\mathbf{b}_0\cdot\left(\br(\mathbf{r})\times\bt(\mathbf{r})\right)\right] \;,
\label{eq:kernel}
\end{split}
\end{align}
where $\Btp$ and $\Brm$ are the co-rotating and counter-rotating parts of the Tx- and Rx-field, respectively. The exponential term in eq.~\eqref{eq:kernel} accounts for phase lags associated with subsurface conductivity structures. The unit vectors $\bt$, $\br$ and $\mathbf{b}_0$ in the third row of eq.~\eqref{eq:kernel} account for the relative orientations of transmitter, receiver and Earth's magnetic field, respectively.

Two general excitation schemes exist in SNMR. When the Tx-pulse is generated by an alternating current $\Iac$ at Larmor frequency $\larmor$, the excitation is called on-resonant (OR) excitation \cite[]{weichman2000PRE}. This is the general case for most SNMR applications. More recent approaches use adiabatic fast passage pulses (AP), where the Tx-pulse current $\Iac$ and its corresponding frequency are modulated in a particular manner to provide a more homogeneous excitation in the subsurface \cite[]{tannus1997NiB,grunewald2016G}. In general $\Mperp$ can be determined for both excitation schemes by solving the Bloch equation $d\Mtotal/d t=\Mtotal\times\gammaH\Btotal$ \cite[]{bloch1946PR}. However, in case of OR-excitation one can omit the time consuming calculation of the Bloch equation and employ directly the solution in the rotating frame of reference \cite[e.g][]{hertrich2008PiNMRS}
\begin{equation}
\Mperp = \sin\left(-\gamma q \left|\Btp(\mathbf{r})\right|\right) = \sin\left(\Theta_{\mathrm{T}}\right)\;,
\label{eq:Mperp}
\end{equation}
where the flip angle $\Theta_{\mathrm{T}}$ describes the orientation of the magnetization after the Tx-pulse. In case of AP-excitation, the effective Tx-field $\Beff$ is modified by a time dependent extra field due to the frequency modulation \cite[e.g.][]{grunewald2016G}. Therefore, we solve the corresponding Bloch equation $d\Mtotal/d t=\Mtotal\times\gammaH\Beff$ to obtain $\Mperp$ and use it in eq.~\eqref{eq:kernel} for the kernel calculation. In section~\ref{sec:theoPPramp} and appendix~\ref{ssec:modeling}, we describe the numerical implementation of the Bloch equation to study PP switch-off ramps. From a numerical perspective there is no difference in using the Bloch equation to determine the magnetization for either the PP- or Tx-pulses.

\begin{figure}
	\centering
	\includegraphics[width=1\linewidth]{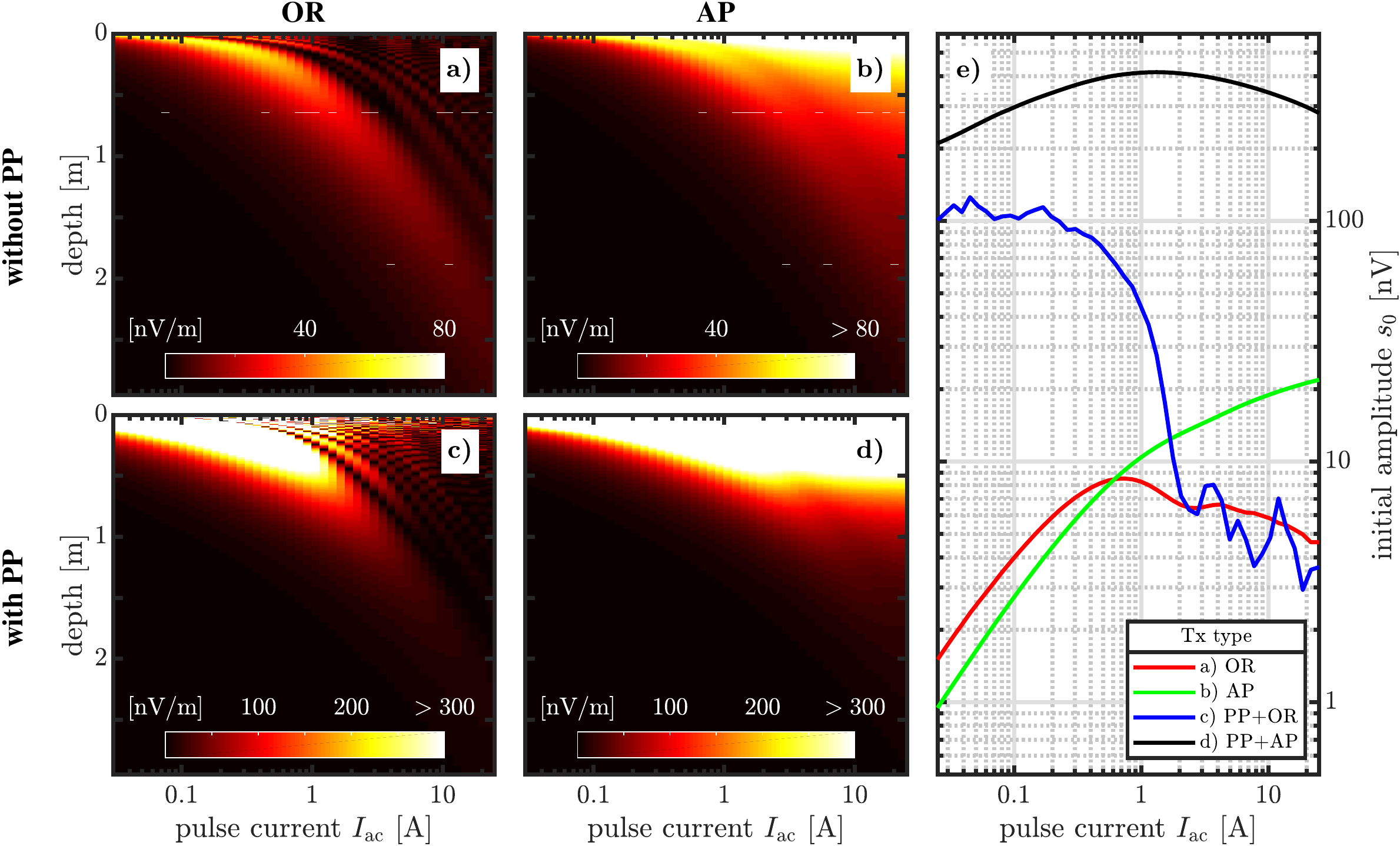}
	\caption{(\textbf{a-d}) sensitivity kernels for two different Tx-excitation pulses -- on-resonant OR (left column) and adiabatic half passage AP (right column) either without pre-polarization (top row) or with pre-polarization (bottom row); Note that the color scale for the PP kernels differs from the kernels without PP; (\textbf{e}) sounding curves calculated from the kernels shown in (\textbf{a-d}) for a homogeneous half space with \SI{30}{\percent} water content.}
	\label{fig:fig2}
\end{figure}

Before studying the effect of PP on SNMR measurements and for introductory reasons, we first compare the effect of OR- and AP-excitation without additional PP. Figure~\ref{fig:fig2}a+b exemplarily shows the corresponding sensitivity kernels and Fig.~\ref{fig:fig2}e the associated sounding curves (red and green lines). The kernel calculations were performed with MRSMatlab \cite[]{mueller-petke2016G} where the magnetic fields are discretized in cylindrical coordinates with number of radial segments $\mathrm{nr}=71$, number of angular segments $\mathrm{n\varphi}=361$ and number of depth layers $\mathrm{nz}=192$. For both kernels a \SI{100}{\ohm\meter} half space and a $d=\SI{2}{\meter}$ coincident circular loop layout (Tx: 1 turn; Rx: 10 turns) are used. The Earth's magnetic field is set to $\Berda=\SI{48}{\micro\tesla}$ ($\larmorf=\SI{-2043.7}{\hertz}$) and the pulse length for both pulses is \SI{40}{\milli\second}. In the AP-case, we employ a so-called adiabatic half-passage pulse \cite[][]{tannus1997NiB} that sweeps from an offset $\Delta f=\SI{-200}{\hertz}$ to the Larmor frequency $\larmorf$ with a hyperbolic tangent. The amplitude modulation for the AP-pulse is given implicitly according to a typical surface NMR coil response with a quality factor of $Q=30$ \cite[]{grunewald2016G}. The pulse current $\Iac$ in Fig.~\ref{fig:fig2} refers to the constant AC current in the OR-case and to the maximum current at the end of the Tx-pulse in the AP-case. The OR-kernel (Fig.~\ref{fig:fig2}a) shows the typical banded sensitivity structure, where an increase in $\Iac$ yields a corresponding deeper depth focus. Here, the deepest band represents the first perfect excitation, i.e. a flip angle of \ang{90}. In contrast to this, the AP-kernel (Fig.~\ref{fig:fig2}b) exhibits a more homogeneous sensitivity structure that has its maximum close to the surface. The sensitivity structure of both kernels also gets reflected in the corresponding sounding curves. Both sounding curves are calculated for a constant water content of \SI{30}{\percent}. For smaller pulse currents $\Iac<\SI{0.5}{\ampere}$, the OR sounding curve (red) has larger amplitudes compared to the AP-case (green). For larger $\Iac$ and due to the increasing homogeneous excitation with depth, the AP-amplitudes increase up to a factor of \num{4} compared to the OR-values at similar $\Iac$. For a more detailed discussion on the implementation of AP-excitation in SNMR and its particular characteristics we refer the reader to \cite{grunewald2016G}.

\section{SNMR-PP with perfect pre-polarization switch-off}
\label{sec:theoPPideal}

To further increase the net magnetic moment $\Mnull$, and thereby increase the detectable signal amplitude, it is possible to temporarily apply a strong pre-polarizing magnetic field $\Bpp$ prior to the actual excitation pulse \cite[][]{packard1954PR}. During the presence of the PP-field $\Bpp$, the spins will align with the resulting field $\Btotal=\Berd+\Bpp$ and hence, the magnetic moment $\Mnull$ will increase to the spatially varying $\Mpp(\mathbf{r})\propto\left|\Berd+\Bpp\right|$ \cite[]{melton1971RSI,planinsic1994JMRA,callaghan1997RoSI,pasquale2014VZJ}. Switching off the polarizing field $\Bpp$ adiabatically causes the spins to reorient towards $\Berd$ and at the same time $\Mpp$ exponentially decreases towards $\Mnull$ with relaxation time constant $T_1$ (cf.~eq.~\ref{eq:bloch}). In case of an adiabatic switch-off, $\Mpp$ is perfectly aligned with $\Berd$ after $\Bpp$ has vanished and $\Mnulla$ in eq.~\ref{eq:kernel} can be replaced by $\Mppa$ to calculate the forward kernel. The adiabatic condition is satisfied if the frequency $\gamma\left\vert\Btotal(t)\right\vert$ remains much larger than the rate of reorientation of $\Btotal$ from $\Bpp$ towards $\Berd$ \cite[]{melton1995JoMRSA,melton2002JoMR}
\begin{equation}
\gamma\left\vert\Btotal(t)\right\vert \gg \frac{d\alpha}{dt}\;,
\label{eq:adiabatic}
\end{equation}
where $\alpha$ is the angle between $\Btotal$ and $\Berd$ (cf.~Fig.~\ref{fig:figA1}a), for the entire duration of the switch-off. To get the maximum yield out of PP, two things need to be considered when switching off the polarizing field $\Bpp$. First, the adiabatic condition itself (eq.~\ref{eq:adiabatic}) has to be satisfied, meaning that the switch-off has to be slow enough, so that $\Mpp$ is ``locked" to $\Btotal$ as it decreases in amplitude and reorients to $\Berd$ \cite[]{melton1995JoMRSA,melton2002JoMR}. If the switch-off is too fast, most of the magnetization will not follow the reorienting $\Btotal$ and will not be aligned in parallel to $\Berd$ at the end of the switch-off. Consequently, these magnetization components do not get coherently excited by a subsequent Tx-pulse and hence, do not contribute to the recorded NMR signal. Second, the switch-off has to be fast enough (much faster than $T_1$) to get the maximum gain out of the increased $\Mpp$. If the switch-off is not much faster than $T_1$, then a considerable amount of magnetization will decay due to $T_1$-relaxation before the actual Tx-pulse is applied. The critical part when switching off $\Bpp$ is at the end of the switch-off when $\Bpp$ becomes small and the precession frequency of $\Mtotal$ around $\Berd$ is small compared to the rate of change of reorientation of $\Btotal$. This reorientation rate needs to slow down in order to maintain the inequality in eq.~\ref{eq:adiabatic}.

Due to technical limitations of current available SNMR-PP devices and depending on the length of the Tx-pulse, it can take about \SI{50}{\milli\second} after the PP switch-off until the actual NMR signal is recorded (cf.~Fig.~\ref{fig:fig1}). Therefore, it is undesirable to have a long PP switch-off (long in respect to the relatively short $T_1$) that adds magnetization losses due to $T_1$-relaxation. Reported switch-off times for recent SNMR-PP devices are about \SI{1}{\milli\second} \cite[]{lin2017GJI,lin2018RoSI,costabel2019NSG} and are therefore about two to three orders of magnitude smaller than typical $T_1$-values for saturated porous media.

\begin{figure}
	\centering
	\includegraphics[width=0.9\linewidth]{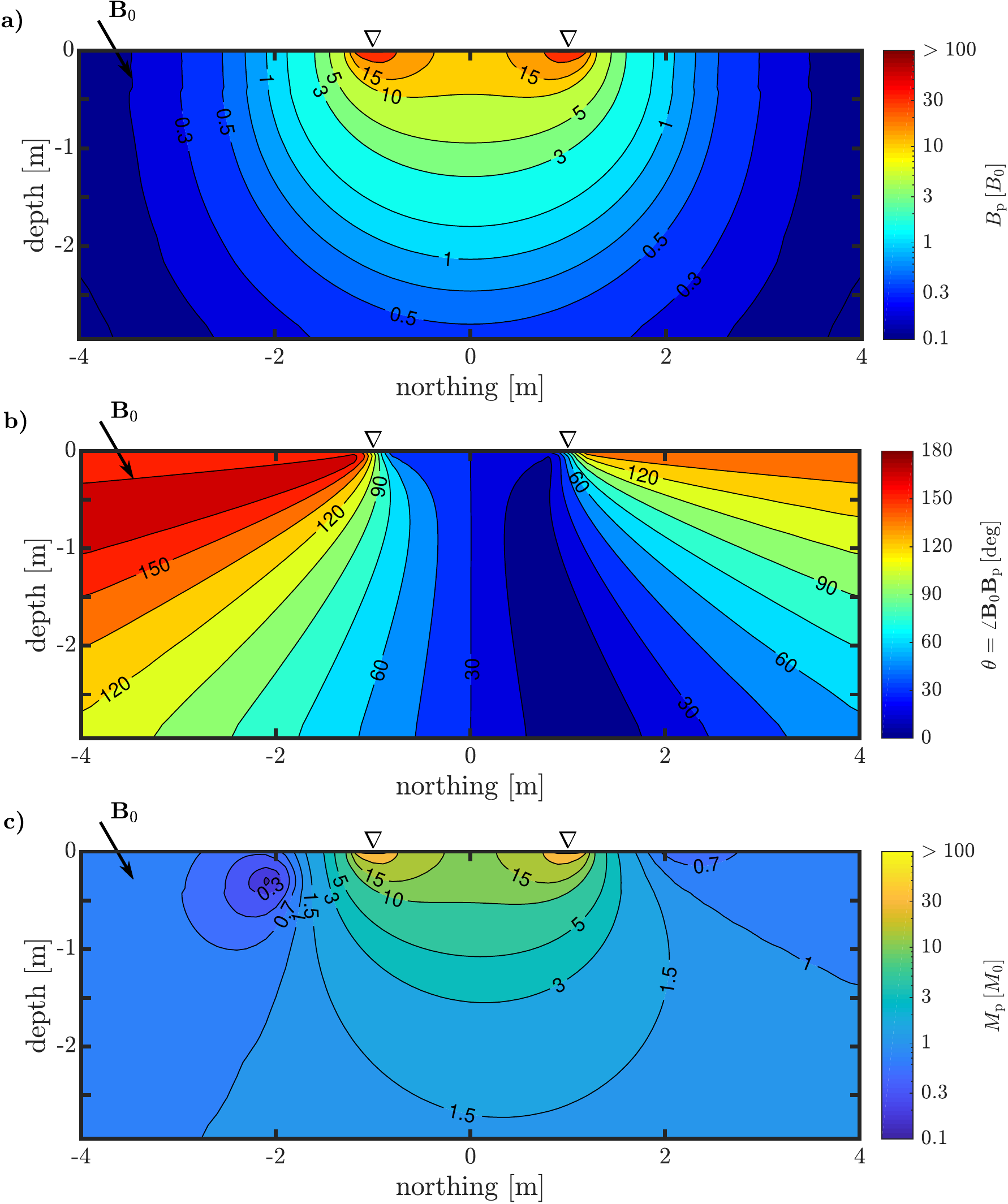}
	\caption{(\textbf{a}) amplitudes $\Bppa=\Bppn$ of a PP-field generated by a circular PP-loop with $d=\SI{2}{\meter}$ and effective DC of $\Ipp=\SI{1000}{\ampere}$; (\textbf{b}) angle $\theta$ between $\Berd$ and $\Bpp$; (\textbf{c}) corresponding amplitudes $\Mppa=\Mppn$ of the amplified magnetization after a perfectly adiabatic switch-off of the $\Bpp$-field shown in (a); all sub panels are 2D slices out of a 3D domain and are oriented S$\rightarrow$N; the vertical triangles in all sub panels indicate the PP-loop position.}
	\label{fig:fig3}
\end{figure}

Figure~\ref{fig:fig3}a exemplarily shows the amplitudes $\Bppa=\Bppn$ in units of the Earth's magnetic field $\Berda$ for a central slice (oriented S$\rightarrow$N) out of a three-dimensional domain. The corresponding PP-field is generated by a \SI{2}{\meter} circular PP-loop with $\Ipp=\SI{1000}{\ampere}$ energizing DC. Only in the very close vicinity of the PP-loop (vertical triangles), $\Bppa$ exceeds values of \num{100} (dark red colors). With increasing distance, $\Bppa$ rapidly decreases to \num{1} (light blue colors) at about \SI{2}{\meter} away from the loop. Because we employ an inclination of \ang{60} for the Earth's magnetic field $\Berd$ and due to the inhomogeneity of the PP-field, there is a spatially varying distribution of angles $\theta$ that describe the relative orientation of $\Bpp$ in relation to $\Berd$ (Fig.~\ref{fig:fig3}b). In the case of perfect adiabatic switch-off and neglecting $T_1$-relaxation, the PP-field in Fig.~\ref{fig:fig3}a+b generates an enhanced magnetization $\Mppa$ as shown in Fig.~\ref{fig:fig3}c. Due to the inclination of $\Berd$ with respect to the PP-loop, the spatial distribution of $\Mppa$ is not symmetric in the given 2D-plane. The maximum enhancement exceeds \num{100} close to the PP-loop (yellow) and is even smaller than \num{1}, at greater distances from the PP-loop (dark blue colors), especially to the south.

To visualize the effect of SNMR-PP, we plot two sensitivity kernels in Fig.~\ref{fig:fig2}c+d (bottom row) and their corresponding sounding curves in Fig.~\ref{fig:fig2}e. The model parameters are identical to the cases without PP and the applied PP-field corresponds to the one shown in Fig.~\ref{fig:fig3}a. Note that with PP, the color bar values in Fig.~\ref{fig:fig2}c+d are larger compared to the cases without PP. For visualization purposes we combined all values larger than \SI{300}{\nano\volt\per\meter} into the color white. For a few points very close to the surface, the sensitivity exceeds \SI{30000}{\nano\volt\per\meter} for the cases with PP.

For both SNMR-PP-cases, the sounding curve amplitudes for small $\Iac$ are about two orders of magnitude larger compared to the cases without PP. For the PP+OR-case (blue) the amplitudes quickly decrease for $\Iac>\SI{0.25}{\ampere}$ and reach values comparable to the ones without PP at about $\Iac=\SI{2.5}{\ampere}$. The fluctuations of the PP+OR sounding curve at larger $\Iac$ compared to the OR-case are notable and attributed to the strong fluctuations of the sensitivity kernel very close to the surface that get traced through even for larger $\Iac$. These fluctuations were already reported by \cite{pasquale2014VZJ} and are due to the particular loop layout. When spins get excited in regions where the $\Bpp$-field is large (close to the loop), the flip angle $\Theta_{\mathrm{T}}$ can get larger than \ang{180} and the amplitude contribution of those regions is negative, hence the oscillating kernel. To suppress these oscillations, the authors suggested to decrease the size of the Tx/Rx-loops by a factor of three compared to the size of the PP-loop. However, as this would also decrease the maximum possible enhancement and penetration depth, we have chosen to employ the aforementioned coincident loop layout. The PP+AP sounding curve (black) shows a different behavior. For small $\Iac$ values the amplification is even higher than for the PP+OR-case (different to the cases without PP). Similarly to the AP-case (green), the amplitudes of the PP+AP signal initially increase up to about $\Iac=\SI{1}{\ampere}$. Although the amplitudes slightly decrease again for larger $\Iac$, they are still more than one order of magnitude higher compared to the AP-case without PP. Due to the much more homogeneous AP-excitation the undesirable kernel oscillations have also vanished.

Both SNMR-PP-kernels show a high sensitivity at the shallow subsurface (bright yellow and white colors) due to the large $\Bpp$-field magnitudes close to the PP-loop and the strong decreasing effect of the $\Bpp$-field with increasing depth (cf.~also Fig.~\ref{fig:fig3}a). This basic comparison demonstrates the advantages of SNMR-PP for shallow subsurface measurements in terms of signal enhancement, especially when combined with AP-excitation.

\section{SNMR-PP with imperfect pre-polarization switch-off}
\label{sec:theoPPramp}

Now, we want to study the effect of an imperfect PP switch-off on the enhancement of the final NMR signal. But before doing so, we compare different switch-off ramps and how the particular ramp shape and ramp time $\tauramp$ influence the switch-off characteristics. In order to reliably quantify a spatially varying subsurface water content distribution it is essential to determine the actual magnetization $\Mpp$ after the switch-off. Considering the practical applicability of SNMR-PP devices, we need to evaluate to what extent a particular switch-off ramp influences the recorded NMR signal. Here, our objective is not to propose or develop a particular switch-off ramp but rather to familiarize the reader with the consequences that may arise when using one of the presented ramps.

To evaluate different PP switch-off ramps, we need to study the full NMR spin dynamics of the magnetization $\Mpp$ during the switch-off of the $\Bpp$-field. To this end, we developed the software BLOCHUS (\textbf{Bloch} \textbf{U}niversal \textbf{S}imulator), a set of Matlab\textsuperscript{TM} tools including a graphical user interface (GUI) that implements the governing equation of motion, the Bloch equation \cite[]{bloch1946PR}, in a laboratory frame of reference
\begin{align}
\begin{split}
\frac{d\Mpp}{dt} = \gamma\Mpp\times\left[\Berd+\Bpp\right] &- \frac{M_{\mathrm{p,x}}\mathbf{e}_{\mathrm{x}}+M_{\mathrm{p,y}}\mathbf{e}_{\mathrm{y}}}{T_2}\\&-\frac{\left(M_{\mathrm{p,z}}-M_{0}\right)\mathbf{e}_{\mathrm{z}}}{T_1}\:,
\label{eq:bloch}
\end{split}
\end{align}
with longitudinal and transversal relaxation times $T_1$ and $T_2$. By controlling the amplitude of the PP-current $\Ipp(t)$ during switch-off, any arbitrary ramp shape and its corresponding effect on $\Bpp$, and hence $\Mpp$, can be modeled. Internally, BLOCHUS solves eq.~\eqref{eq:bloch} with Matlab's ``ode45" routine, which employs a fifth-order Runge-Kutta method with adaptive time stepping \cite[]{shampine1997matlab}. To use the extensive modeling capabilities of MRSmatlab \cite[]{mueller-petke2016G}, BLOCHUS also contains an interface to the kernel calculation routines provided by MRSmatlab. As mentioned above and without loss of generality, eq.~\eqref{eq:bloch} can also be used to model the magnetization $\Mtotal$ during a Tx-pulse (either OR or AP) and therewith implicitly consider effects like relaxation during pulse \cite[]{grombacher2017G}. In appendix~\ref{ssec:modeling}, we provide two benchmark scenarios to validate our implementation of eq.~\eqref{eq:bloch}.

\subsection{Parameter Study}
\label{ssec:parameter}

\begin{figure}
	\centering
	\includegraphics[width=\linewidth]{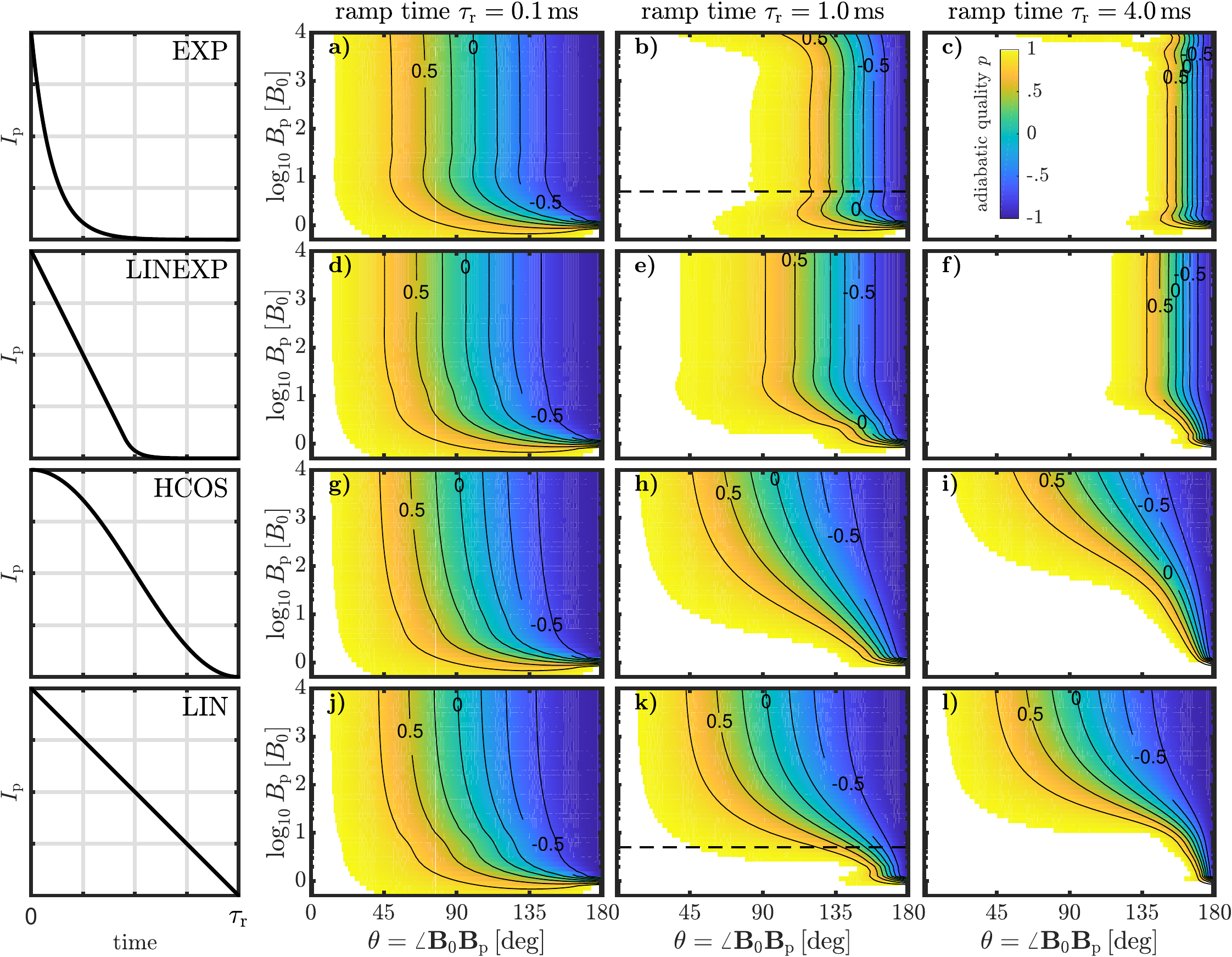}
	\caption{Adiabatic quality $p$ as a function of $\Bppa$ and angle $\theta$ for four different ramp shapes (rows) and three different switch-off times $\tauramp$ (columns); $p=1$ (white) indicates perfect adiabatic switch-off.}
	\label{fig:fig4}
\end{figure}

As we have shown in Fig.~\ref{fig:fig3}a+b, the amplitudes $\Bppa$ and the relative orientation $\theta$ strongly vary in magnitude and spatial extent. Therefore, and depending on the shape and duration of the switch-off ramp, the PP switch-off is not perfect in every point of the subsurface volume. To quantify this, we determine the ``adiabatic quality'' $p$ as a function of amplitude $\Bppa$ and angle $\theta$ for various switch-off ramps (cf.~Fig.~\ref{fig:fig4}). Herein, $p=\mathrm{proj}_{\Berd}\Mpp$ is the normalized projection of the magnetization $\Mpp$ onto the final magnetic field $\Btotal=\Berd$ after the switch-off \cite[]{conradi2017JoMR}. The case $p=1$ describes the ideal case, i.e. $\Mpp$ and $\Berd$ are perfectly parallel after switch-off and the full magnitude of $\Mpp$ can get excited by a subsequent Tx-pulse. The values $p=0$ and $p=-1$ cover the perpendicular and anti-parallel case, respectively and would lead to a decreased Tx-excitation. In the following, we will use $p$ as an effective measure to evaluate the quality of the PP switch-off.

We exemplarily employ four different ramp shapes:
\begin{itemize}
	\item \myexp{} -- exponential \cite[]{lin2017GJI,lin2018RoSI}
	\item \mylinexp{} -- linear \& exponential \cite[e.g][]{conradi2017JoMR}
	\item \myhalfcos{} -- half cosine \cite[]{costabel2019NSG}
	\item \mylin{} -- linear \cite[e.g][]{melton1995JoMRSA}
\end{itemize}
which have been published before and are partly in use in current SNMR-PP devices. Each ramp shape was calculated for three different switch-off times $\tauramp$ of \SIlist{0.1;1;4}{\milli\second} where the ramp time $\tauramp$ is defined as the total time span that is used to ramp the $\Bpp$-field down to zero. Note that similar to a real SNMR-PP measurement, $\tauramp$ is fixed for all combinations of $\Bppa$ and $\theta$ within the corresponding column of Fig.~\ref{fig:fig4}. For all calculations we neglected $T_1$- and $T_2$-relaxation.

With the chosen set of ramp shapes and ramp times, we cover a reasonable range of parameters for technical applicable SNMR-PP measurements. A ramp time of $\tauramp=\SI{1}{\milli\second}$ is reported in \cite{lin2018RoSI} and \cite{costabel2019NSG} and the two other ramp times serve as ``boundary" cases to evaluate the dependence of $p$ onto $\tauramp$. We note that the short ramp time of $\tauramp=\SI{0.1}{\milli\second}$ might be challenging to implement in a real SNMR-PP device, especially when considering the strong currents involved. The ramp shapes are chosen to cover a range of shapes that show distinct differences at early and late times, respectively. Considering the adiabatic criterion in eq.~\ref{eq:adiabatic}, obviously using the \myexp{}-ramp seems to be a generally favorable choice as it releases a lot of energy over a short period of time (and early) while it significantly slows down at later ramp times. On the contrary, the \myhalfcos{}-ramp releases the energy much more slowly and uniformly and will therefore violate eq.~\ref{eq:adiabatic} over a larger period of time. However, as this ramp is also in practical use, we want to evaluate its effect on the switch-off quality. The other two shapes are the benchmark cases from appendix~\ref{ssec:modeling}, where for the cases presented here the switch-over field strength for the \mylinexp{}-ramp is fixed to $\Bstar=\Bppa/10$. In Fig.~\ref{fig:fig4} the adiabatic quality $p$ is color coded from white ($p=1$, perfect adiabatic switch-off) over yellow, green ($p=0$) to blue ($p=-1$).

By basic inspection of the $p$-distributions in Fig.~\ref{fig:fig4}, several features can be recognized. First and in line with the adiabatic criterion in eq.~\ref{eq:adiabatic}, for each ramp shape (rows in Fig.~\ref{fig:fig4}) the global adiabatic quality increases with increasing ramp time $\tauramp$. By ``global" we refer to the whole range of $\Bppa$ and $\theta$ values considered here. However, the increase of the global adiabatic quality itself with increasing $\tauramp$ is significantly different for the individual ramp shapes. For the case of $\tauramp=\SI{0.1}{\milli\second}$ there seems to be no significant difference between the individual ramp shapes. In this particular case, all ramp shapes show an equally poor performance when compared to corresponding results for longer ramp times. Within the relevant PP-parameters, a switch-off time of \SI{0.1}{\milli\second} is simply too fast to allow for an adiabatic reorientation if the initial angle between $\Bpp$ and $\Berd$ is already larger then $\theta\approxeq\ang{45}$. For a ramp time of $\tauramp=\SI{1}{\milli\second}$ the global adiabatic quality $p$ shows a distinct dependence on the ramp shape. For the \myexp{}-ramp, the switch-off is perfect up to an angle of $\theta\approxeq\ang{115}$ (Fig.~\ref{fig:fig4}b). For the \mylinexp{}, the \myhalfcos{} and the \mylin{} ramps this onset of low adiabatic quality happens continuously ``earlier" at smaller angles $\theta$ (Fig.~\ref{fig:fig4}e,h,k). Furthermore, while for the \myexp{}-ramp the distribution of $p$ is rather independent on the magnitude of $\Bppa$, the dependence on $\Bppa$ is continuously increasing for the \mylinexp{}, the \myhalfcos{} and the \mylin{} ramps. For the latter ones, the global adiabatic quality increases with decreasing $\Bppa$. The results for $\tauramp=\SI{4}{\milli\second}$ are along the same line of arguments as for the results for $\tauramp=\SI{1}{\milli\second}$. The main difference is that a high adiabatic quality can be achieved over a larger range of $\theta$ values. These findings seem to confirm that for a large range of $\Bppa$ and $\theta$ values, the \myexp{}-ramp outperforms the \mylinexp{}, \myhalfcos{} and \mylin{} ramps and therefore should yield the maximum PP-effect in a real SNMR-PP measurement.

\begin{figure}
	\centering
	\includegraphics[width=1\linewidth]{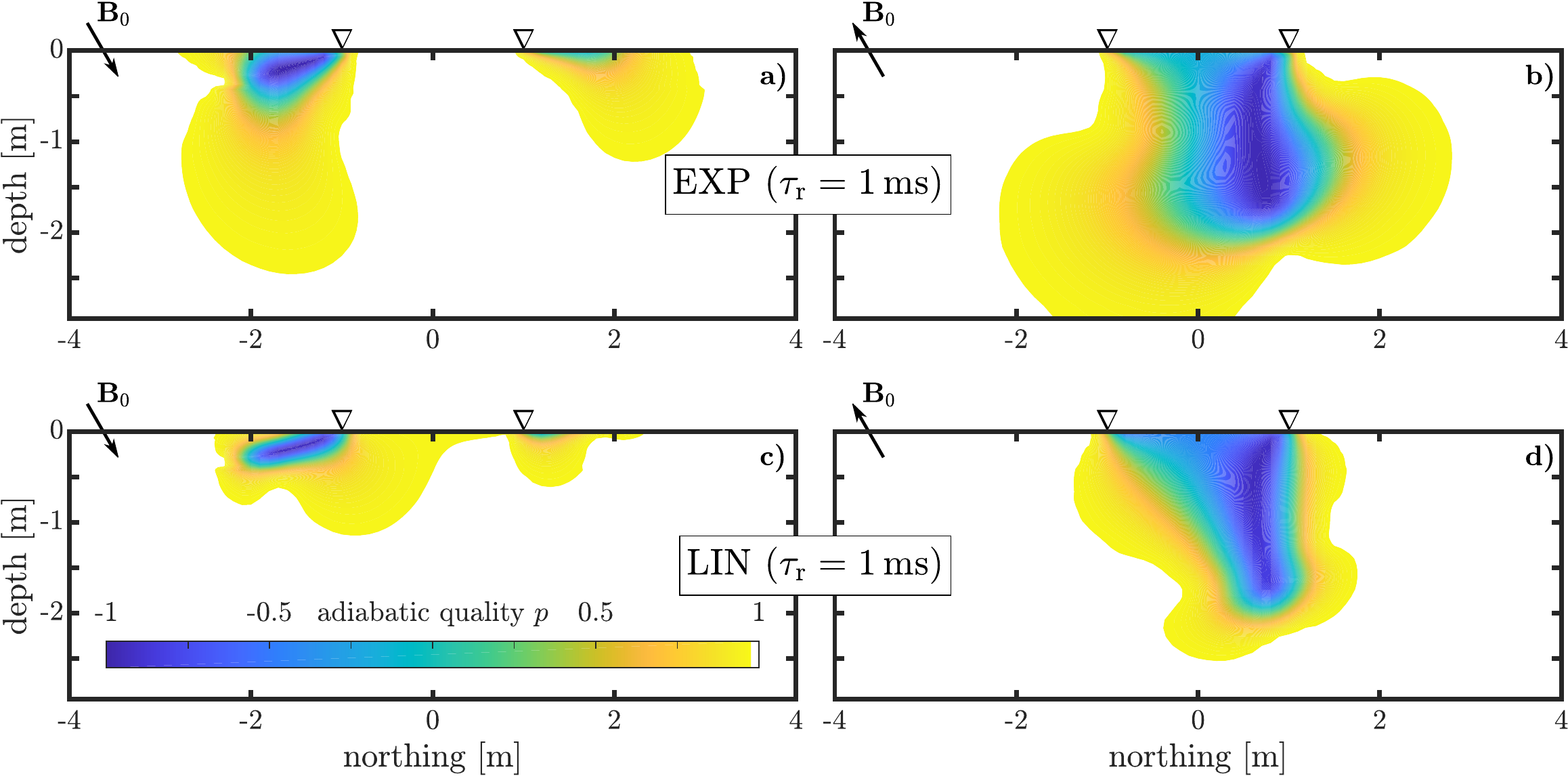}
	\caption{Spatial distribution of adiabatic quality $p$ for the \myexp{} (\textbf{a+b}) and \mylin{} (\textbf{c+d}) switch-off ramps with a switch-off time of $\tauramp=\SI{1}{\milli\second}$; In (\textbf{a+c}) the inclination is \ang{60} and in (\textbf{b+d}) the inclination is \ang{-60}; all sub panels are 2D slices out of a 3D domain and are oriented N$\rightarrow$S; the vertical triangles in all sub panels indicate the PP-loop position.}
	\label{fig:fig5}
\end{figure}

To further evaluate these results in regard of a real SNMR-PP measurement, we calculate the spatial distribution of adiabatic quality $p$ for the \myexp{} and \mylin{}-ramps with $\tauramp=\SI{1}{\milli\second}$ as they differ most significantly from each other in terms of global $p$ at this particular $\tauramp$ (cf.~Fig.~\ref{fig:fig4}). We use the amplitudes $\Bppa$ and angles $\theta$ from the PP-field presented in Fig.~\ref{fig:fig3}a+b and determine $\Mpp$ and therewith also $p$ based on the Bloch equation eq.~\eqref{eq:bloch}. Figure~\ref{fig:fig5} shows the corresponding central slices of the 3D distribution of $p$ when either an \myexp{}-ramp (Fig.~\ref{fig:fig5}a+b) or \mylin{}-ramp (Fig.~\ref{fig:fig5}c+d) is used. For each ramp, we determine $p$ for two different inclinations: \ang{60} (Fig.~\ref{fig:fig5}a+c) and \ang{-60} (Fig.~\ref{fig:fig5}b+d). This not only illustrates the spatial variability of the switch-off quality depending on the relative orientation of $\Berd$ and $\Bpp$, but once more demonstrates that it might be necessary to account for these effects already in the forward calculation.

Despite the general better global performance of the \myexp{}-ramp compared to the \mylin{}-ramp (cf.~Fig.~\ref{fig:fig4}), this ramp yields a larger volume of low $p$-values (Fig.~\ref{fig:fig5}). By a detailed evaluation of Figs.~\ref{fig:fig3},\ref{fig:fig4}+\ref{fig:fig5} it becomes apparent that for values smaller than about $\Bppa<5$ (dashed line in Fig.~\ref{fig:fig4}b+k), the \mylin{}-ramp (Fig.~\ref{fig:fig4}k) shows indeed a higher adiabatic quality over a wider range of angles $\theta$ compared to the \myexp{}-ramp (Fig.~\ref{fig:fig4}b). Whereas for amplitudes $\Bppa>10$, the adiabatic quality of the \mylin{}-ramp significantly decreases. This can be seen e.g. in Fig.~\ref{fig:fig5}c below and inside the loop where the $p$-values are smaller than \num{1}. For the case of \ang{-60}  inclination (Fig.~\ref{fig:fig5}b+d) the situation is similar to the one described before. The total volume of reduced $p$-values is larger for the \myexp{}-ramp than for the \mylin{}-ramp. Additionally, due to the reversed inclination a large amount of magnetization is oriented antiparallel below the center of the loop. In this case a substantial amount of magnetization would not contribute to the NMR signal. A simple way to deal with this kind of situation is to reverse the direction of the PP-current \cite[]{conradi2017JoMR} and therewith establish the $p$-distribution as shown for the \ang{60} inclination case (Fig.~\ref{fig:fig5}a+c). However, depending on the particular inclination of the Earth's magnetic field, there will be points in the subsurface where $p<1$ and therefore also the usable magnetization $\Mppa$ will be reduced correspondingly.

\subsection{Effect of imperfect switch-off on SNMR-PP sounding curves}
\label{ssec:soundingcurves}

\begin{figure}
	\centering
	\includegraphics[width=\linewidth]{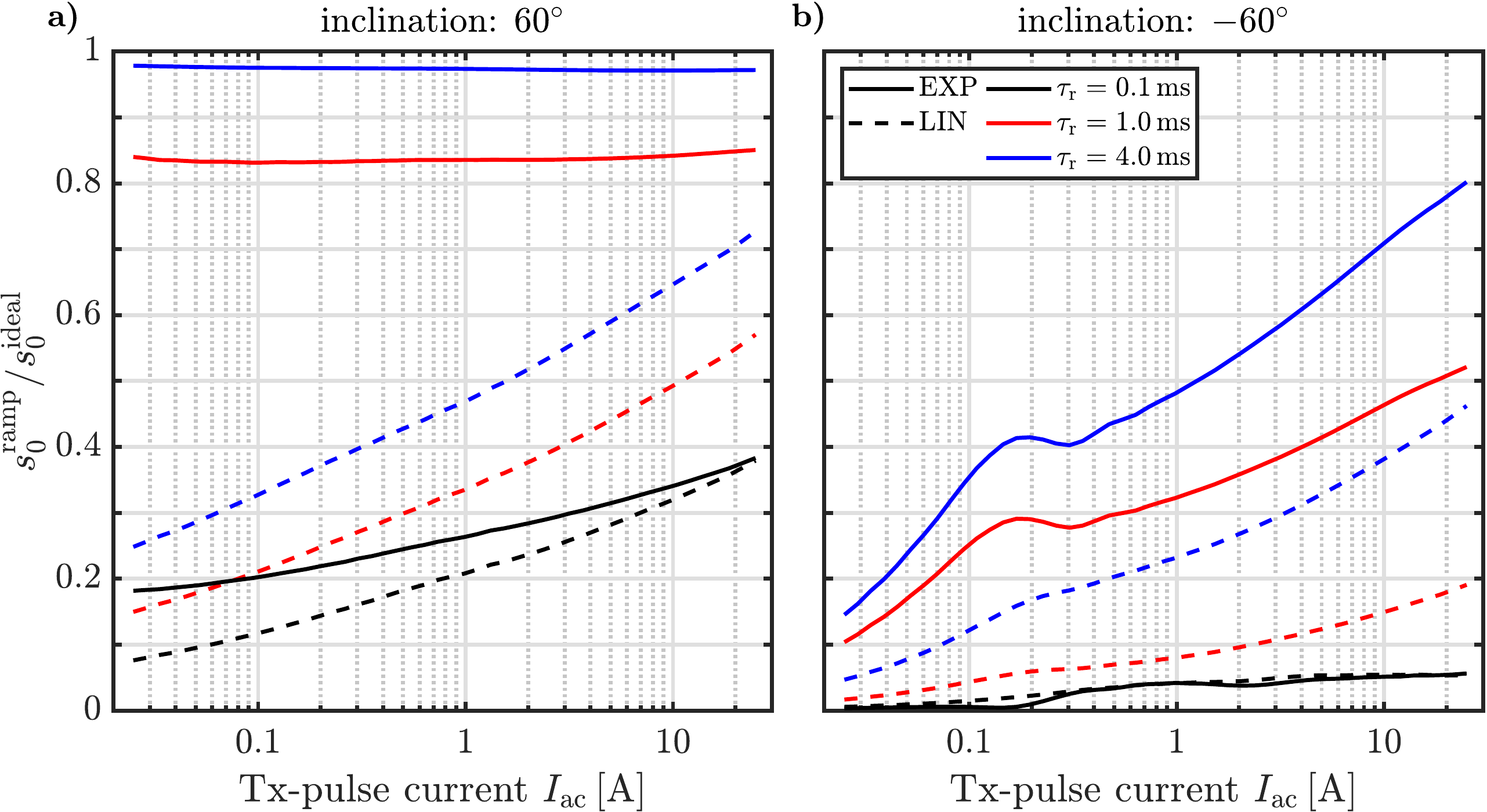}
	\caption{Normalized sounding curves as a function of Tx-pulse current $\Iac$ for the \myexp{} (solid) and \mylin{} (dashed) switch-off ramps for three different switch-off times $\tauramp$ (colors); all sounding curves are normalized by the PP+AP sounding curve with perfect adiabatic switch-off (cf.~Fig.~\ref{fig:fig2}b); In (\textbf{a}) the inclination is \ang{60} and in (\textbf{b}) the inclination is \ang{-60}.}
	\label{fig:fig6}
\end{figure}

So far, we have shown that an imperfect, i.e. non-adiabatic, switch-off has a considerable effect on the spatial distribution of $p$ and that it strongly depends on the particular PP switch-off ramp. However, a meaningful insight into this effect can only be gained by quantifying it in terms of signal strength of the SNMR-PP sounding curve. To this end, we use the SNMR-PP scheme with AP-excitation pulses as introduced in section~\ref{sec:theoPPideal}. This yields the largest signal enhancement regarding our measurement configuration and additionally also suppresses the unwanted kernel oscillations if instead used with OR-excitation (cf.~Fig.~\ref{fig:fig2}). In order to calculate sounding curves similar to the ones in Fig.~\ref{fig:fig2}e, we replace the equilibrium magnetization $\Mnulla$ in eq.~\eqref{eq:kernel} with the component of $\Mpp$ that is parallel to $\Berd$ after the switch-off. Remember that only magnetization components parallel to $\Berd$ get coherently excited by a subsequent Tx-pulse and result in a measurable NMR signal. All presented sounding curves employ the same modeling parameters as described in section~\ref{sec:theoPPideal} for the PP+AP case. In practice this means that for every sounding curve, we use a particular fixed ramp shape and ramp time combination and calculate the three-dimensional distribution of enhanced magnetization $\Mpp$ that is established after the PP switch-off. The magnetization $\Mpp$ gets excited by an AP Tx-pulse and the corresponding voltage response is recorded in a Rx-loop at the surface. The initial values of this recorded NMR-signal are plotted as a function of final Tx-pulse current $\Iac$ in Fig.~\ref{fig:fig6}. In addition and to focus on the signal loss due to the particular ramp parameters, we normalize all curves with the PP+AP-case with theoretically perfect adiabatic switch-off (Fig.~\ref{fig:fig2}b black curve). This means the closer the amplitude of a curve in Fig.~\ref{fig:fig6} is to $s_{0}^{\mathrm{ramp}}/s_{0}^{\mathrm{ideal}}=1$, the more adiabatic is the switch-off and the better is the signal enhancement compared to the ideal case.

Figure~\ref{fig:fig6} shows sounding curves corresponding to the \myexp{} and \mylin{} ramps with solid and dashed lines, respectively. Color coded are the different switch-off ramp times of the corresponding ramp. Let us first consider the case for \ang{60} inclination (Fig.~\ref{fig:fig6}a). For a switch-off time of $\tauramp=\SI{1}{\milli\second}$ (red) the \myexp{}-ramp sounding curve retains about \SI{83}{\percent} signal amplitude compared to the perfect case over the entire Tx-pulse current range. For the \mylin{}-ramp on the other hand, the amplitude loss is much more dramatic with only \SI{15}{\percent} (low $\Iac$) to \SI{55}{\percent} (high $\Iac$) compared to the perfect case. The large difference between the two sounding curves partly originates from the individual switch-off performance of the corresponding ramp at large PP-field strengths present inside / below the PP-loop (cf.~Fig.~\ref{fig:fig4}b+k) and partly from the high sensitivity at the shallow subsurface for this particular loop layout and AP-excitation (cf.~Fig.~\ref{fig:fig2}a bottom right). Although the volume having reduced $p$-values is larger for the \myexp{}-ramp (cf.~Fig.~\ref{fig:fig5}), the resulting sounding curve deviates much less from the perfect case compared to the \mylin{}-ramp. This is due to the better performance of the \myexp{}-ramp for larger $\Bppa$ amplitudes at smaller angles $\theta$ (cf.~Fig.~\ref{fig:fig4}) within the sensitive region of the Tx/Rx-loops. For a very short ramp time of $\tauramp=\SI{0.1}{\milli\second}$ (black), both sounding curves retain less then \SI{40}{\percent} of the signal amplitude over the entire range of Tx-pulse currents. For a longer ramp time of $\tauramp=\SI{4}{\milli\second}$ (blue) the \myexp{}-ramp retains more than \SI{98}{\percent} of the signal amplitude of the perfect case. Even for this quite long switch-off time, the \mylin{}-ramp achieves only an average value of about \SI{50}{\percent}. For the \mylin{}-ramp, the signal loss at small Tx-pulse currents (close to the surface and hence closer to the PP-loop) is stronger than for larger Tx-pulse currents, which is again attributed to the poor performance at large values of $\Bppa$ (cf.~Fig.~\ref{fig:fig4}) compared to the \myexp{}-ramps. In this comparison we left out the \mylinexp{} and \myhalfcos{} ramps for visibility reasons and because we want to put focus on the qualitative differences. The sounding curves for both of these ramps plot between the sounding curves for the \myexp{} and \mylin{} ramps for every ramp time we considered. The results for \ang{-60} inclination impressively show how severe the effect of the relative orientation of the magnetic fields is. Even for a long switch-off time of $\tauramp=\SI{4}{\milli\second}$ the signal amplitude of the \myexp{}-ramp at small $\Iac$ is reduced to less than \SI{30}{\percent}. Furthermore, all sounding curves show a strong dependence on $\Iac$, where for larger $\Iac$ the performance generally increases.

It is not surprising that ramps which fulfill the adiabatic condition much better during switch-off (\myexp{}, \mylinexp{}) perform much better than e.g. the \myhalfcos{} and \mylin{} ramps. The observed differences between the individual ramps might be significantly decreased for loop layouts where the Tx/Rx-loops are placed in a particular manner, so that their sensitive region is further away from the PP-loop and the relative orientations between $\Berd$ and $\Bpp$ are more favorable. However, this always comes at the cost of reduced magnetization enhancement and penetration depth \cite[]{pasquale2014VZJ}. The cases presented here, should therefore be regarded as a kind of worst case scenarios in terms of underestimating the NMR signal amplitude if the particular PP switch-off ramp is not considered in the forward calculation. More importantly, any signal loss, as presented in Fig.~\ref{fig:fig6}, is directly misinterpreted as reduced water content when performing the inversion of the SNMR-PP data.
 
\section{Summary \& Conclusions}
\label{sec:conclusion}

The combined use of pre-polarization (PP) and adiabatic excitation pulses (AP) is a promising approach to significantly enhance the signal quality of SNMR measurements that target shallow subsurface water content distributions. The nowadays available SNMR-PP devices allow investigations down to depths of about \SI{2}{\meter} depending on loop size and available PP-current. A typical SNMR-PP excitation consists of a strong PP-pulse that aligns the enhanced magnetization with the prevailing magnetic field. After the PP-field is switched off adiabatically, the magnetization aligns with the Earth's magnetic field and gets excited by a subsequent Tx-pulse. In theory, the amplified magnetization is properly aligned with the Earth's magnetic field at the end of the PP switch-off. Due to the inhomogeneity of the PP-field and the generally fixed settings of the particularly employed PP switch-off ramp, there are always regions in the subsurface where the PP switch-off is non-adiabatic, i.e. imperfect. The spatial distribution of these imperfect regions mainly depends on the relative orientation between the Earth's magnetic field and the PP-field and therefore of course implicitly on the inclination of the Earth's magnetic field.

To quantify the effect of an imperfect PP switch-off on SNMR-PP signals, we developed a numerical scheme that allows to study arbitrary ramp shapes and incorporates the PP switch-off simulation directly into the well-established SNMR modeling framework MRSmatlab. For parameters typically used in current SNMR-PP applications the enhancement performance varied between \SI{15}{\percent} and \SI{83}{\percent} of the perfect PP-case, depending on the chosen ramp parameters and Tx-pulse current (cf.~Fig.~\ref{fig:fig6}). This is particularly interesting because until now, all SNMR-PP modeling studies assume adiabatic, i.e. perfect, switch-off conditions for the PP-field \cite[]{pasquale2014VZJ,lin2017GJI,lin2018RoSI,lin2019IEEE}. Depending on the severity of the performance loss, this effectively leads to an underestimation of the inverted water content.

The presented modeling framework directly incorporates the PP switch-off into the SNMR-PP forward calculation. This allows to correctly account for the PP switch-off in already available SNMR-PP devices. Furthermore, it enables the user to evaluate and optimize different PP/Tx/Rx-loop layouts and PP/Tx-excitation parameters during the development of a new generation of SNMR-PP devices. Inherently, the performance of any switch-off ramp will depend on the available power electronics of the particular SNMR-PP device. Regarding the maximal possible magnetization enhancement, it seems advantageous to use exponentially shaped ramps that fulfill the adiabatic condition almost during the entire switch-off time. However, in our opinion a generalization on the optimal switch-off ramp shape and time is difficult to make and depends on the particular application. Our future work will focus on the development of optimal PP/Tx/Rx measurement setups and parameters, in order to qualitatively and quantitatively target shallow subsurface water content distributions.

\begin{acknowledgments}
We would like to thank two anonymous reviewers for their comments and suggestions. Furthermore, we would like to thank Stephan Costabel and Eiichi Fukushima for fruitful discussions and constructive comments on the application of SNMR-PP. This work was supported by the German Research Foundation under the grant MU 3318/4-1.
\end{acknowledgments}

\bibliography{literature}

\appendix

\section{Verification of spin dynamics modeling}
\label{ssec:modeling}

To validate our implementation of eq.~\eqref{eq:bloch}, we use two published examples that study the effect of different switch-off characteristics on pre-polarization effectiveness. The first is from \cite{melton1995JoMRSA}, where the authors use linear switch-off ramps with varying switch-off rates, to find a criteria for the so-called sudden passage (the opposite of adiabatic passage). The second example is from \cite{conradi2017JoMR}, where the authors use switch-off ramps that consist of an early linear and a late exponential part. In their work, the authors show how the cross-over field strength $\Bstar$ between these two regimes influences the adiabatic quality of the switch-off, depending on i.e. the initial orientation of $\Berd$ and $\Bpp$.

\begin{figure}
	\centering
	\includegraphics[width=\linewidth]{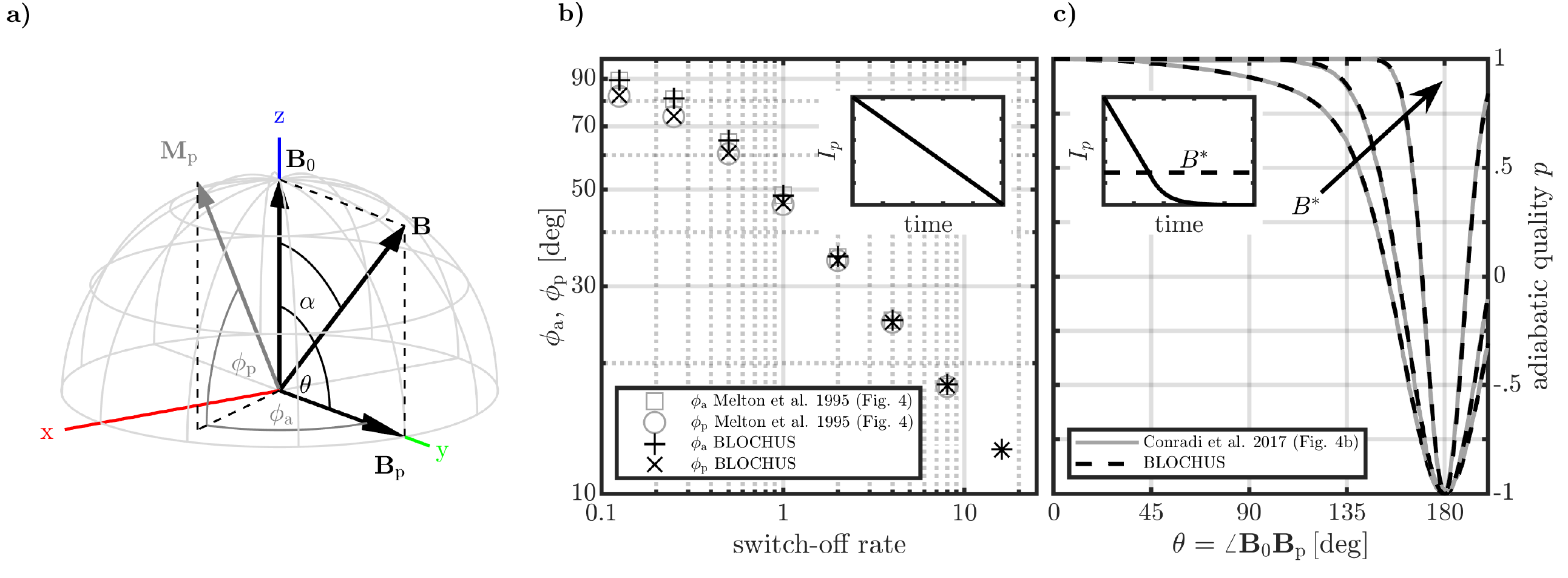}
	\caption{\textbf{a)} Sketch of the geometrical relations used in this work; \textbf{b)} comparison of our implementation (black symbols) to \cite{melton1995JoMRSA} (gray symbols), azimuthal and polar angles $\phi_{\mathrm{a}}$ and $\phi_{\mathrm{p}}$ as a function of switch-off rate (see text); \textbf{c)} comparison of our implementation (black dashed lines) to \cite{conradi2017JoMR} (gray lines), adiabatic quality $p$ as a function of initial angle $\theta$ for three different cross-over field strengths $\Bstar$.}
	\label{fig:figA1}
\end{figure}

A schematic representation of the geometry involved in these two studies and this work is shown in Fig.~\ref{fig:figA1}a. By convention, the primary field $\Berd$ is collinear with the z-axis. The angle $\theta$ describes the orientation between the Earth's magnetic field $\Berd$ and the $\Bpp$-field prior to switch-off. The angle $\alpha$ is the angle between $\Berd$ and $\Btotal=\Berd+\Bpp$ that varies during switch-off as $\Btotal$ moves from its initial orientation towards $\Berd$. The polar and azimuthal angles $\phi_{\mathrm{p}}$ and $\phi_{\mathrm{a}}$ describe the final orientation of $\Mpp$ with regard to $\Bpp$.

In the first example, the initial angle is fixed at $\theta=\SI{90}{\degree}$ and the amplitude of the PP-field is $\Bppa=100\cdot\Berda$ with $\Berda=\SI{50}{\micro\tesla}$. Fig.~\ref{fig:figA1}b shows the angles $\phi_{\mathrm{p}}$ and $\phi_{\mathrm{a}}$ as a function of the dimensionless switch-off rate $100/\larmor/\tauramp$. \cite{melton1995JoMRSA} found that the larger the switch-off rate is (shorter ramp time), the larger is the deviation of the final orientation of $\Mpp$ with regard to $\Berd$, and hence, the switch-off is no longer adiabatic. The gray symbols in Fig.~\ref{fig:figA1}b refer to the results of \cite{melton1995JoMRSA} and the black symbols are modeled with BLOCHUS. We see an excellent agreement over the whole range of switch-off rates.

In the second example the $\Bpp$-field amplitude is $\Bppa=50\cdot\Berda$ with $\Berda=\SI{50}{\micro\tesla}$ and the switch-off ramp time is fixed to $\tauramp=\SI{10}{\milli\second}$. We plot the adiabatic quality $p$ as a function of initial angle $\theta\in\left[\SI{0}{\degree},\SI{200}{\degree}\right]$ (Fig.~\ref{fig:figA1}c). The gray curves in Fig.~\ref{fig:figA1}c refer to the results of \cite{conradi2017JoMR} for different cross-over field strengths $\Bstar$. They show that the larger the cross-over field strength is, the better is the adiabatic quality over a wider range of initial angles $\theta$. For this particular set of parameters, this means that the earlier the switch to the exponential regime happens, the more adiabatic is the switch-off. Again, our results (black dashed curves) show an excellent agreement with the published data. For the details of the two studies, we refer the reader to \cite{melton1995JoMRSA} and \cite{conradi2017JoMR}.

\label{lastpage}

\end{document}